\documentclass{PoSno}

\title{Meson-exchange currents and quasielastic predictions for neutrino-nucleus scattering}

\ShortTitle{Meson-exchange currents}

\author{\speaker{M.B. Barbaro}\\
        Dipartimento di Fisica, Universita' di Torino and INFN, Torino, Italy\\
        E-mail: \email{barbaro@to.infn.it}}

\author{J.E. Amaro\\
        Departamento de F\'{\i}sica At\'omica, Molecular y Nuclear,
and Instituto de F\'{\i}sica Te\'orica y Computacional Carlos I,
Universidad de Granada, Spain\\
        E-mail: \email{amaro@ugr.es}}

\author{J.A. Caballero\\
        Departamento de F\'{\i}sica At\'omica, Molecular y Nuclear,
        Universidad de Sevilla, Spain\\
                E-mail: \email{jac@us.es}}

\author{A. De Pace\\
        Istituto Nazionale di Fisica Nucleare, Sezione di Torino, Italy\\
        E-mail: \email{depace@to.infn.it}}

\author{T.W. Donnelly\\
     Center for Theoretical Physics, Laboratory for Nuclear
  Science and Department of Physics, Massachusetts Institute of Technology,
  Cambridge, MA 02139, USA\\
        E-mail: \email{donnelly@mit.edu}}

\author{G.D. Megias\\
        Departamento de F\'{\i}sica At\'omica, Molecular y Nuclear,
        Universidad de Sevilla, Spain\\
                E-mail: \email{megias@us.es}}

\author{I. Ruiz Simo\\
        Departamento de F\'{\i}sica At\'omica, Molecular y Nuclear,
and Instituto de F\'{\i}sica Te\'orica y Computacional Carlos I,
Universidad de Granada, Spain\\
        E-mail: \email{ruizsig@ugr.es}}

\abstract%
{
  We review some recent progress in the study of electroweak interactions in nuclei within the SuSAv2-MEC model. 
The model has the capability to predict (anti)neutrino scattering observables on different nuclei.  
The theoretical predictions are compared with the recent  T2K $\nu_\mu-^{16}$O data and good agreement is found at all kinematics.  
The results are very similar to those obtained for $\nu_\mu-^{12}$C  scattering, except at low energies, where some differences emerge.  
The role of meson-exchange currents in the two-particle two-hole channel is analyzed in some detail. In particular it is shown that the density dependence of these contributions is different from what is found for the quasielastic response.
  }
  
\FullConference{The 19th International Workshop on Neutrinos from Accelerators-NUFACT2017\\
		25-30 September, 2017\\
		Uppsala University, Uppsala, Sweden}

\begin{document}

Nuclear physics plays a crucial role in the analysis of neutrino
oscillation experiments: nuclear modeling uncertainties in the description of neutrino-nucleus scattering represent the main source of systematic error for long baseline neutrino experiments which aim at precision measurements of neutrino oscillation parameters.

In order to test and constrain nuclear models to be used in these analyses, it is necessary to use informations provided by other experiments, in particular electron-nucleus scattering data.
This is the basis of the SuSA model~\cite{Amaro:2004bs},
which exploits the scaling and superscaling properties exhibited by electron scattering data in order to predict neutrino-nucleus observables.
In its more recent version, SuSAv2~\cite{Gonzalez-Jimenez:2014eqa}, the model
also takes into account the behavior of the responses provided
by the Relativistic Mean Field (RMF): in particluar, the natural enhancement of the transverse electromagnetic response provided by RMF, a genuine dynamical
relativistic effect, is incorporated in the SuSAv2 approach.
However, while the RMF approach works properly at low to intermediate
values of the momentum transfer $q$, where the effects linked to the treatment of the final-state interactions (FSI) are significant, it fails at higher $q$
due to the strong energy-independent scalar and vector RMF potentials, whose effects should instead become less and less important with increasing momentum transfer. In this regime the relativistic
plane-wave impulse approximation (RPWIA) is indeed more appropriate.
Therefore, the SuSAv2 model incorporates both approaches, RMF and RPWIA,
and combines them through a $q$-dependent ``blending'' function that
allows a smooth transition from low/intermediate 
(validity of RMF) to high (RPWIA-based region) $q$-values.

The SuSAv2 predictions for inclusive $(e,e')$ scattering on $^{12}$C  have been presented in \cite{Megias:2016lke}, where they are shown to provide a remarkably good description of the data for very different kinematical situations. In order to perform such comparison the SuSAv2 model has been extended from the the quasielastic (QE) domain to the inelastic region by employing phenomenological fits to the single-nucleon inelastic electromagnetic structure functions.
Furthermore, ingredients beyond the impulse approximation, namely two-particle-two-hole (2p2h) excitations, have been added to the model. These contributions, corresponding to the coupling of the probe to a pair of interacting nucleons and associated to two-body meson exchange currents (MEC), are known to play a very significant role in the ``dip'' region between the QE and $\Delta$ peaks.  In the SuSAv2 approach they are treated  within the Relativistic Fermi Gas (RFG)
model, which allows for an exact and fully relativistic calculation, as required for the extended kinematics involved in neutrino reactions.

\begin{figure}[!htb]
  \begin{center}
    \includegraphics[scale=0.56]{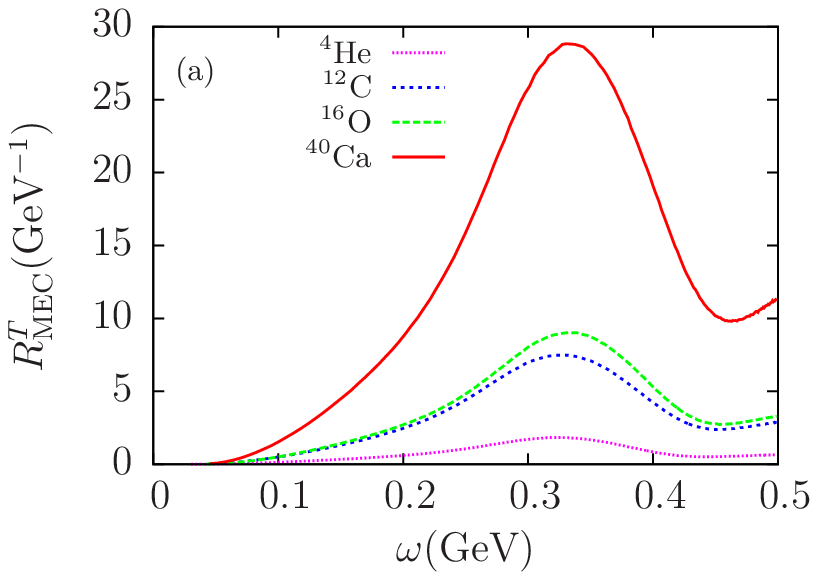}\hspace{0.4cm}\includegraphics[scale=0.56]{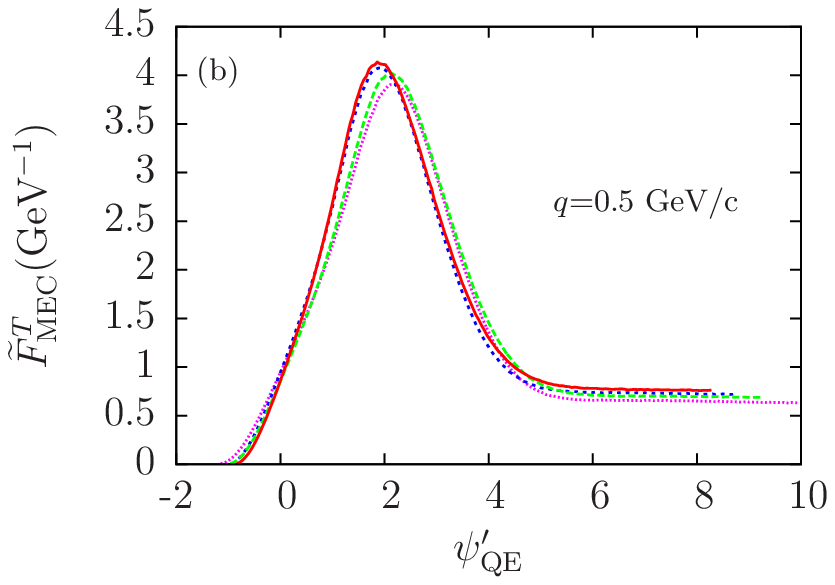}\hspace{0.4cm}\includegraphics[scale=0.56]{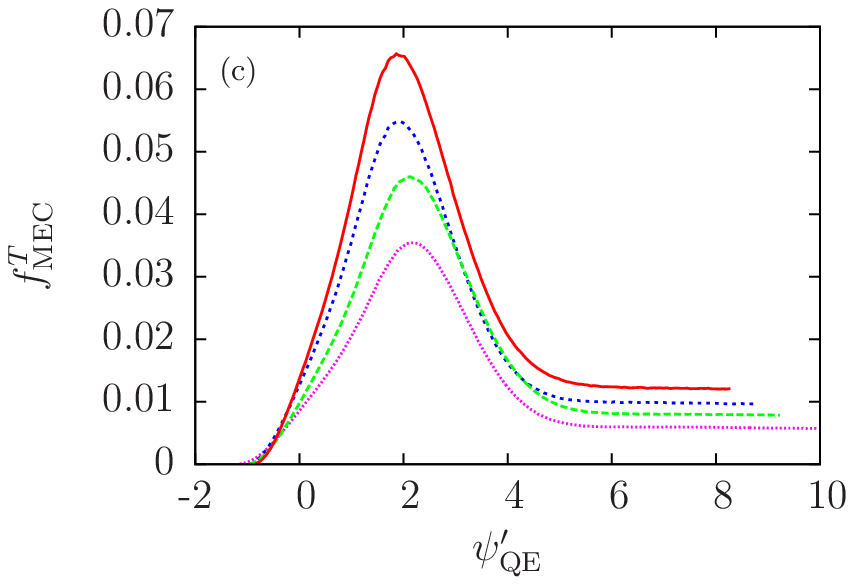}
    \\
    \includegraphics[scale=0.56]{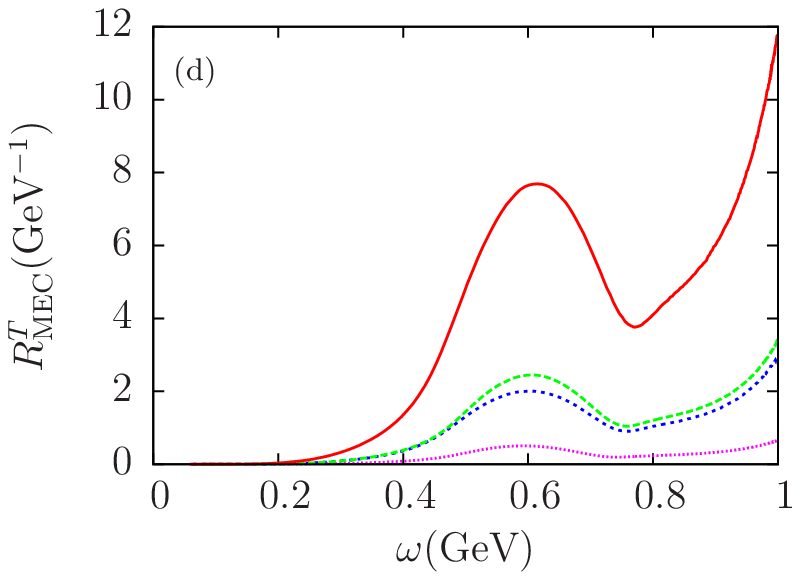}\hspace{0.5cm}\includegraphics[scale=0.56]{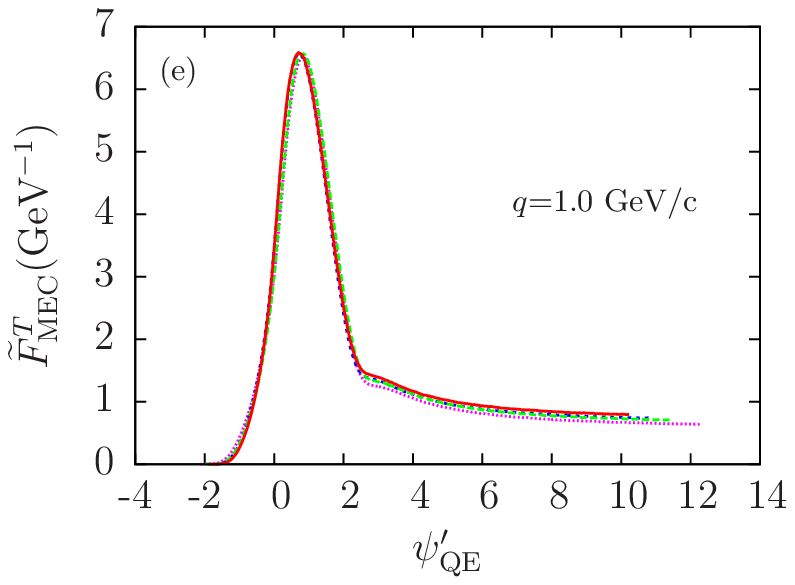}\hspace{0.5cm}\includegraphics[scale=0.56]{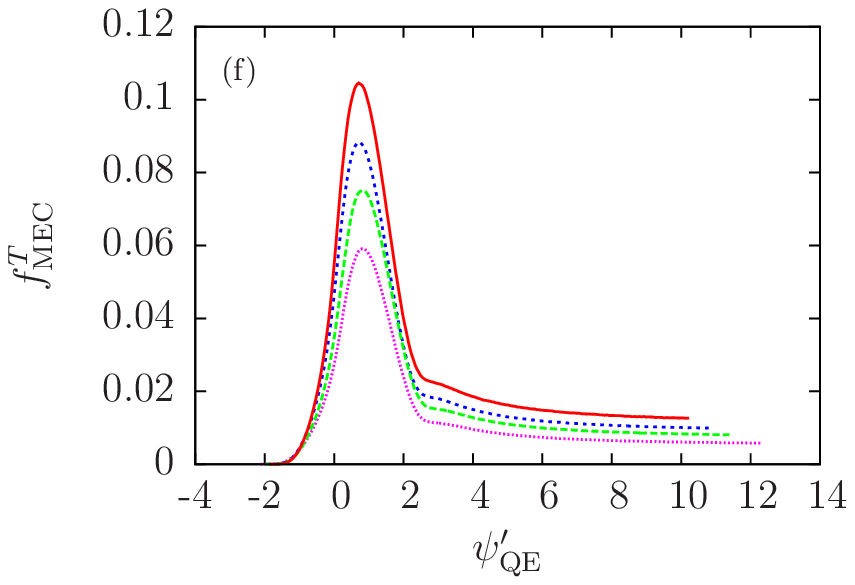}
\\
    \includegraphics[scale=0.56]{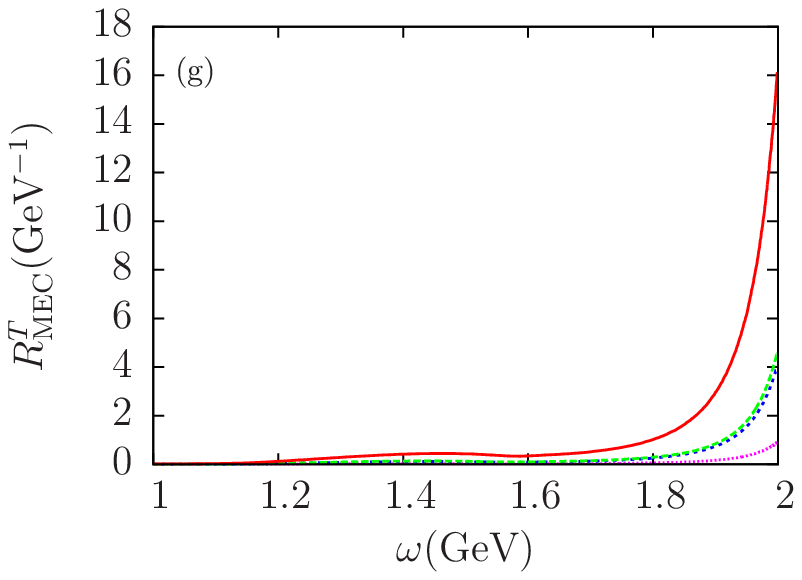}\hspace{0.5cm}\includegraphics[scale=0.56]{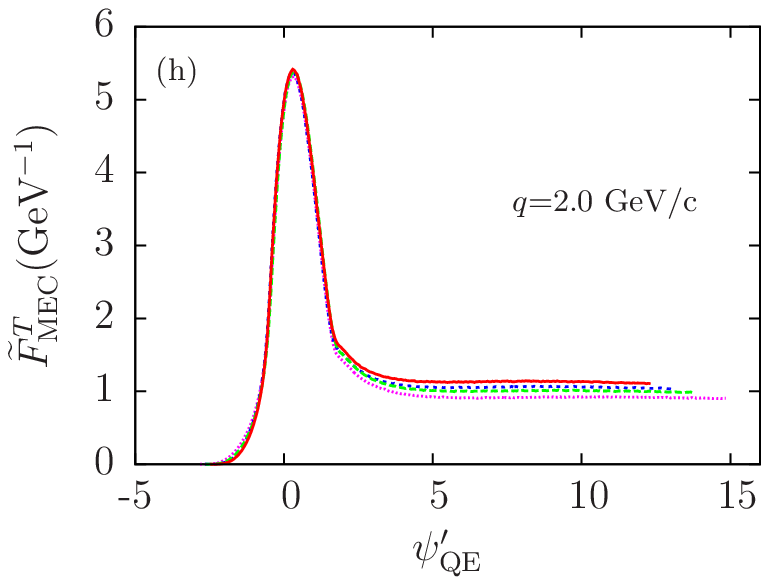}\hspace{0.5cm}\includegraphics[scale=0.56]{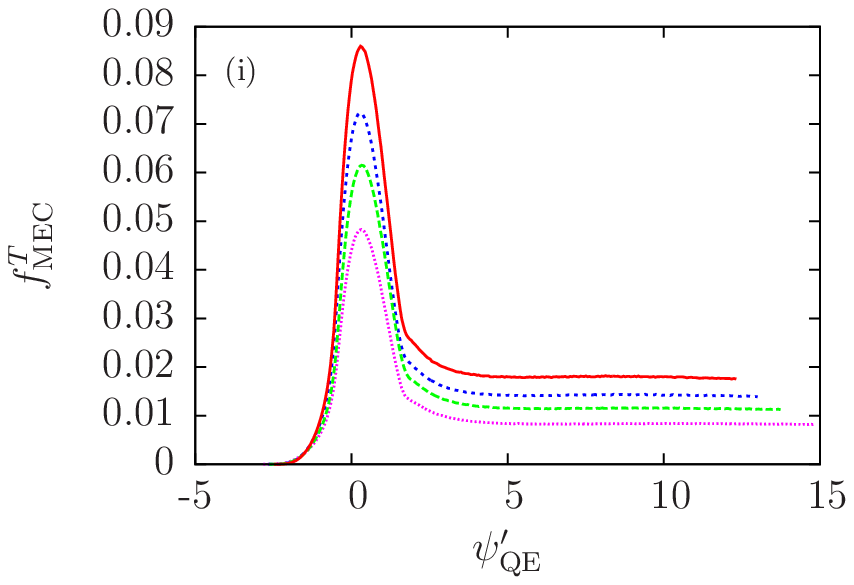}
    \caption{(Color online) The 2p2h MEC response (first column), the corresponding scaled response $\widetilde F^T_{\rm MEC}$ defined by Eq. (\ref{eq:tilf}) (second column) and the superscaling function $f^T_{\rm MEC}$ defined by Eq. (\ref{eq:fsup}) (third column)
      for four nuclei and three values of momentum transfer $q$.
    Figure from Ref.~\cite{Amaro:2017eah}.}
\label{fig:fig8}
\end{center}
\end{figure}

The increasing experimental interest in theoretical predictions for neutrino cross sections on targets other than $^{12}$C, specifically $^{40}$Ar and $^{16}$O, requires the extension of the above calculation performed for carbon to different nuclei.
In this context, the similarities and differences between
charged current (anti)neutrino scattering with no pions in the final
state (the so-called CC0$\pi$ process) on $^{16}$O and $^{12}$C
  have been explored~\cite{Megias:2017cuh}.
The CC0$\pi$ process receives contributions from two different reaction mechanisms: QE scattering and excitation of 2p2h states.
These two mechanisms in general have different dependences on the nuclear
species, namely they scale differently with the nuclear density.
This was shown in Ref.~\cite{Amaro:2017eah} and is illustrated in Fig.~\ref{fig:fig8}, where the transverse electromagnetic MEC response, $R^T_{\rm MEC}$, is plotted versus the energy transfer $\omega$, together with the functions
\begin{equation}
  \widetilde F^T_{\rm MEC}
  \equiv \frac{m_N^2}{k_F^2} \times
\frac{R^T_{\rm MEC}(q,\omega)}{ Z G_{Mp}^2(\tau) + N G_{Mn}^2(\tau)}
  \equiv \frac{m_N^2}{k_F^2} \times F^T_{\rm MEC}(q,\omega)
\label{eq:tilf}
\end{equation}
and
\begin{equation}
  f^T_{\rm MEC}
  \equiv k_F\times
\frac{R^T_{\rm MEC}(q,\omega)}{ Z G_{Mp}^2(\tau) + N G_{Mn}^2(\tau)} 
  \equiv k_F\times F^T_{\rm MEC}(q,\omega)
\label{eq:fsup}
\end{equation}
plotted versus the usual quasielastic scaling variable $\psi^\prime_{\rm QE}$~\cite{Amaro:2004bs} for three values of the momentum transfer $q$ and for the symmetric nuclei $^{4}$He, $^{12}$C, $^{16}$O and $^{40}$Ca. In the above equations $k_F$ is the Fermi momentum, $F^T_{\rm MEC}(q,\omega)$ is the reduced MEC transverse response and $G_{Mp(n)}$ is the proton (neutron) magnetic form factor.  The cases of $^{12}$C and $^{16}$O are clearly relevant for ongoing neutrino oscillation studies, whereas the case of $^{40}$Ca is a symmetric nucleus lying close to the important case of $^{40}$Ar. For comparison, $^{4}$He is also displayed.
 The results show that the reduced 2p2h response $F^T_{\rm MEC}(q,\omega)$ roughly scales as $k_F^2$ when
represented as a function of $\psi^\prime_{\rm QE}$ (second column), {\it i.e.}, the scaled 2p2h MEC response shown there coalesces at the peak into a universal result. This scaling law is even more accurate at the peak of the 2p2h response when a scaling variable $\psi^\prime_{\rm MEC}$, specifically devised for this region, is used (see Ref.~\cite{Amaro:2017eah}). On the other hand the usual second-kind scaling observed in the QE regime, which would require that the function $f^T_{\rm MEC}$ be independent of $k_F$, is clearly violated (third column).

\begin{figure}[!htb]
\begin{minipage}{\textwidth}
  \begin{center}
\includegraphics[width=3.7cm,angle=270]{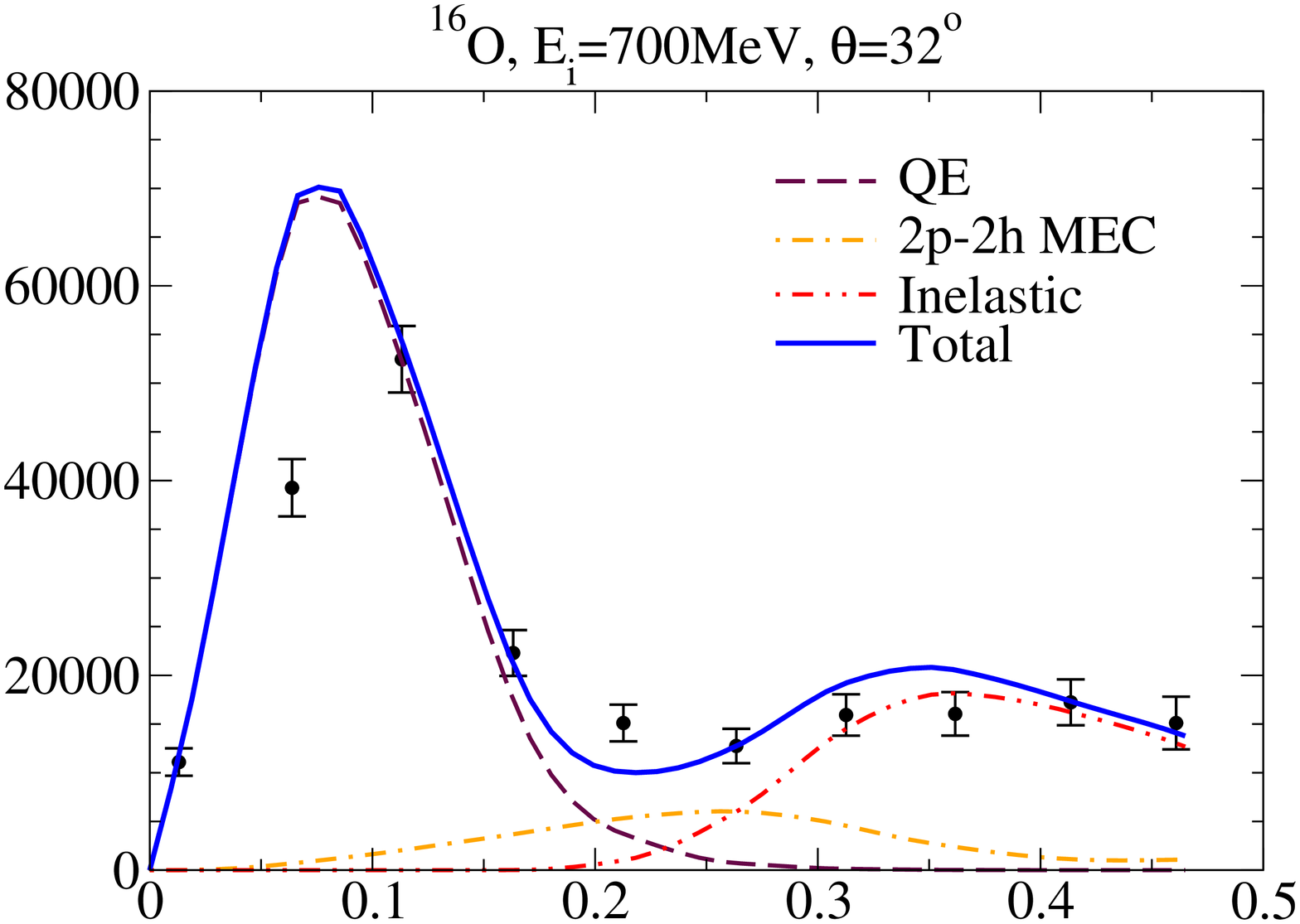}
\includegraphics[width=3.7cm,angle=270]{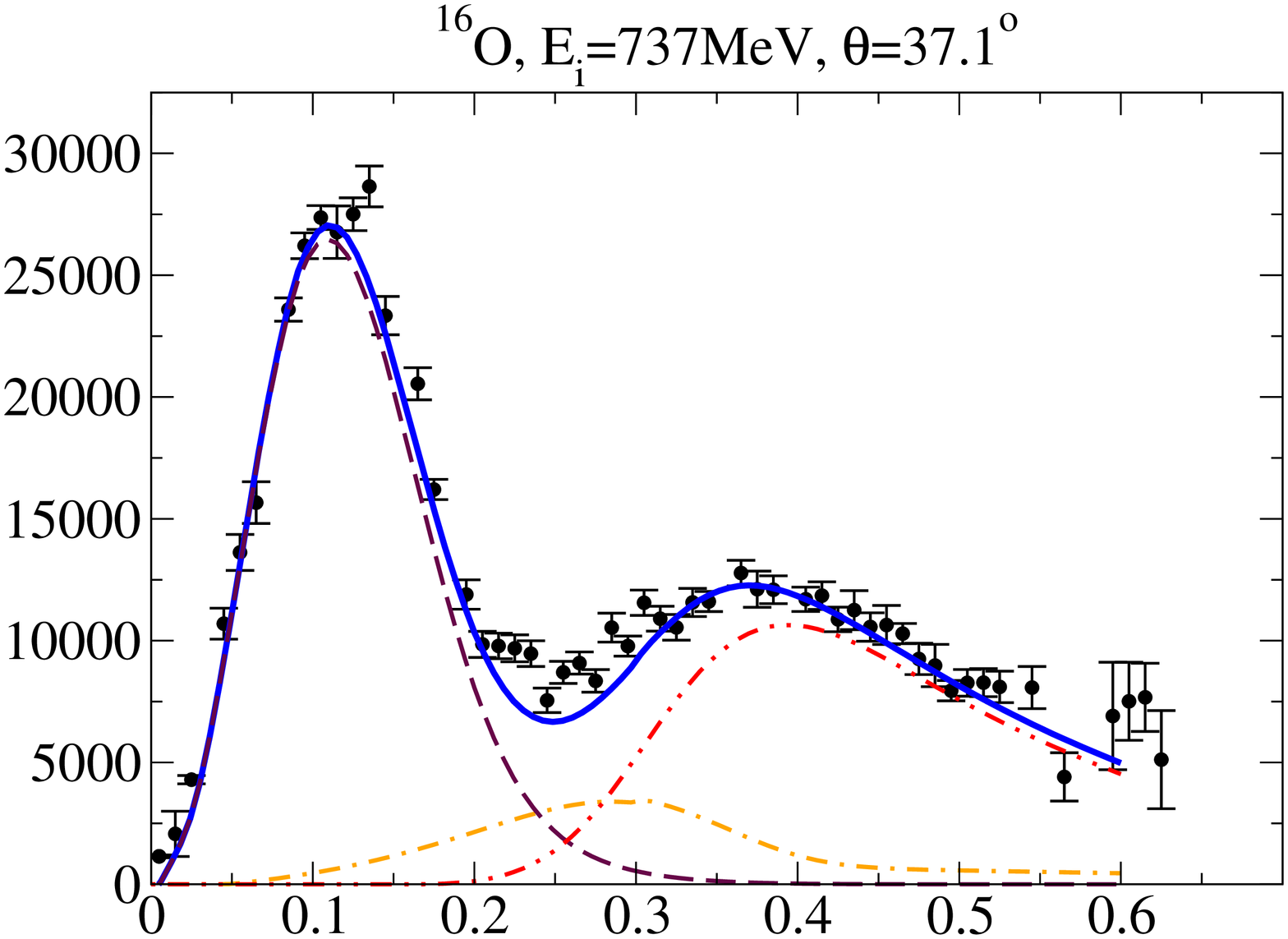}
\includegraphics[width=3.7cm,angle=270]{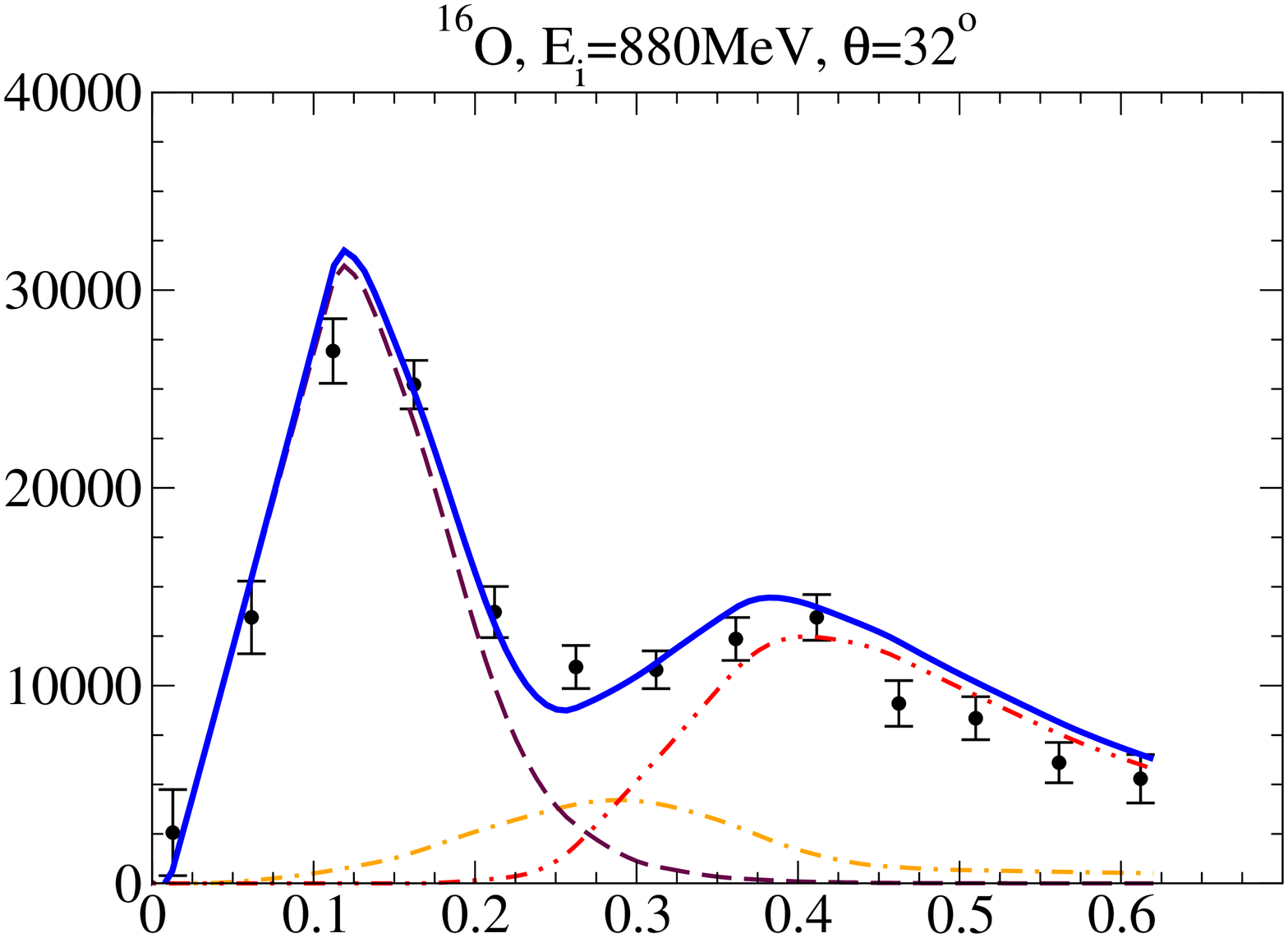}
\\
\includegraphics[width=3.7cm,angle=270]{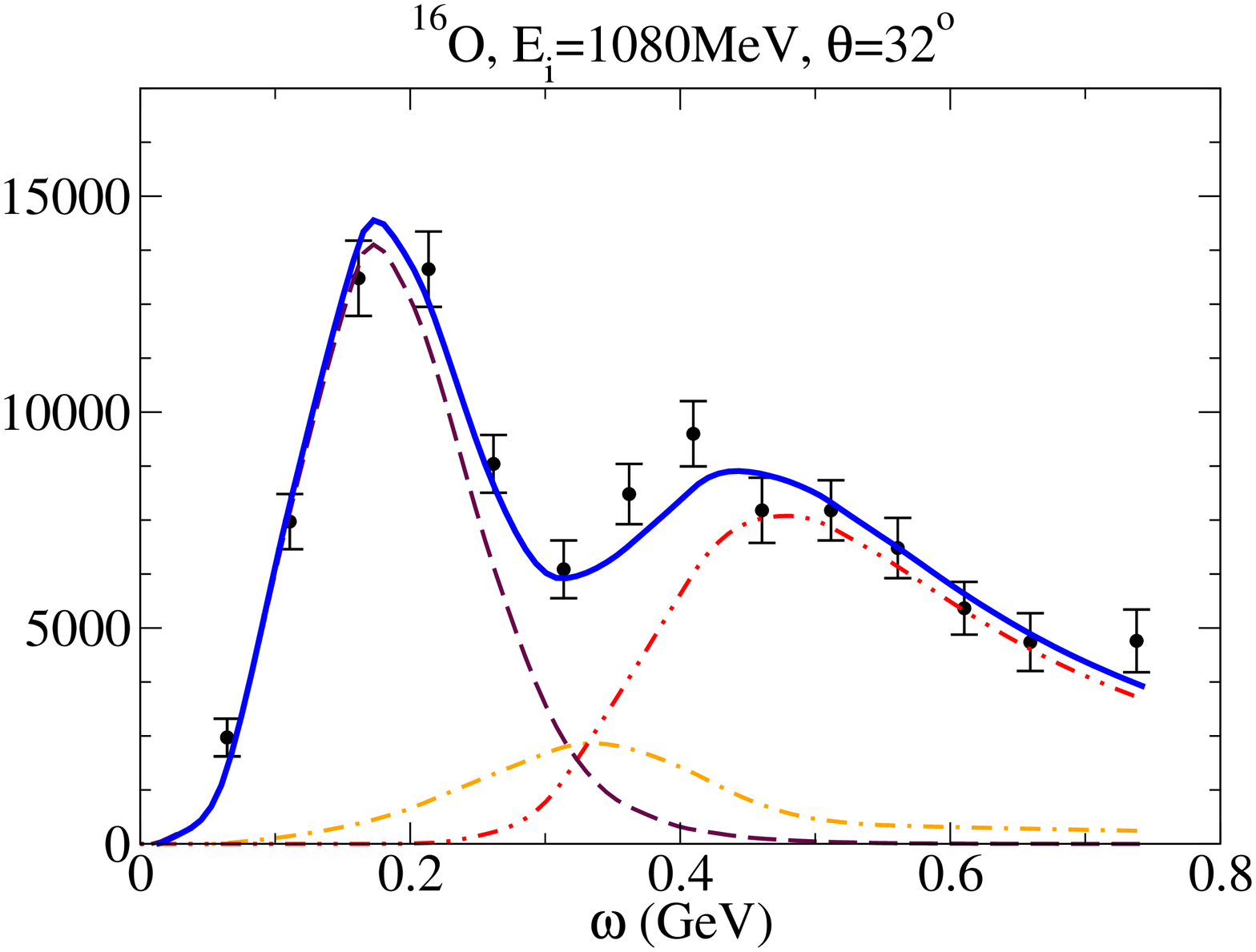}
\includegraphics[width=3.7cm,angle=270]{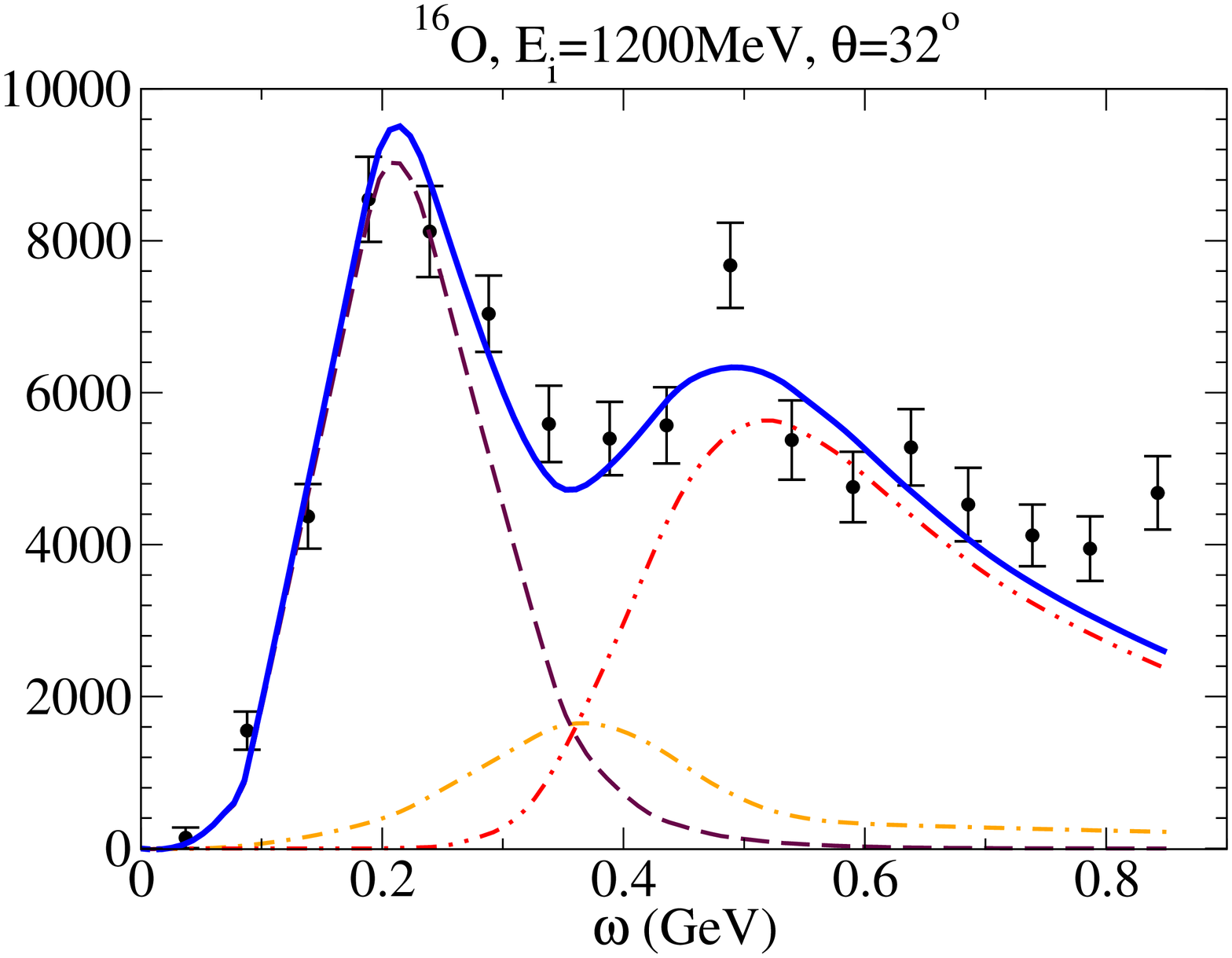}
\includegraphics[width=3.7cm,angle=270]{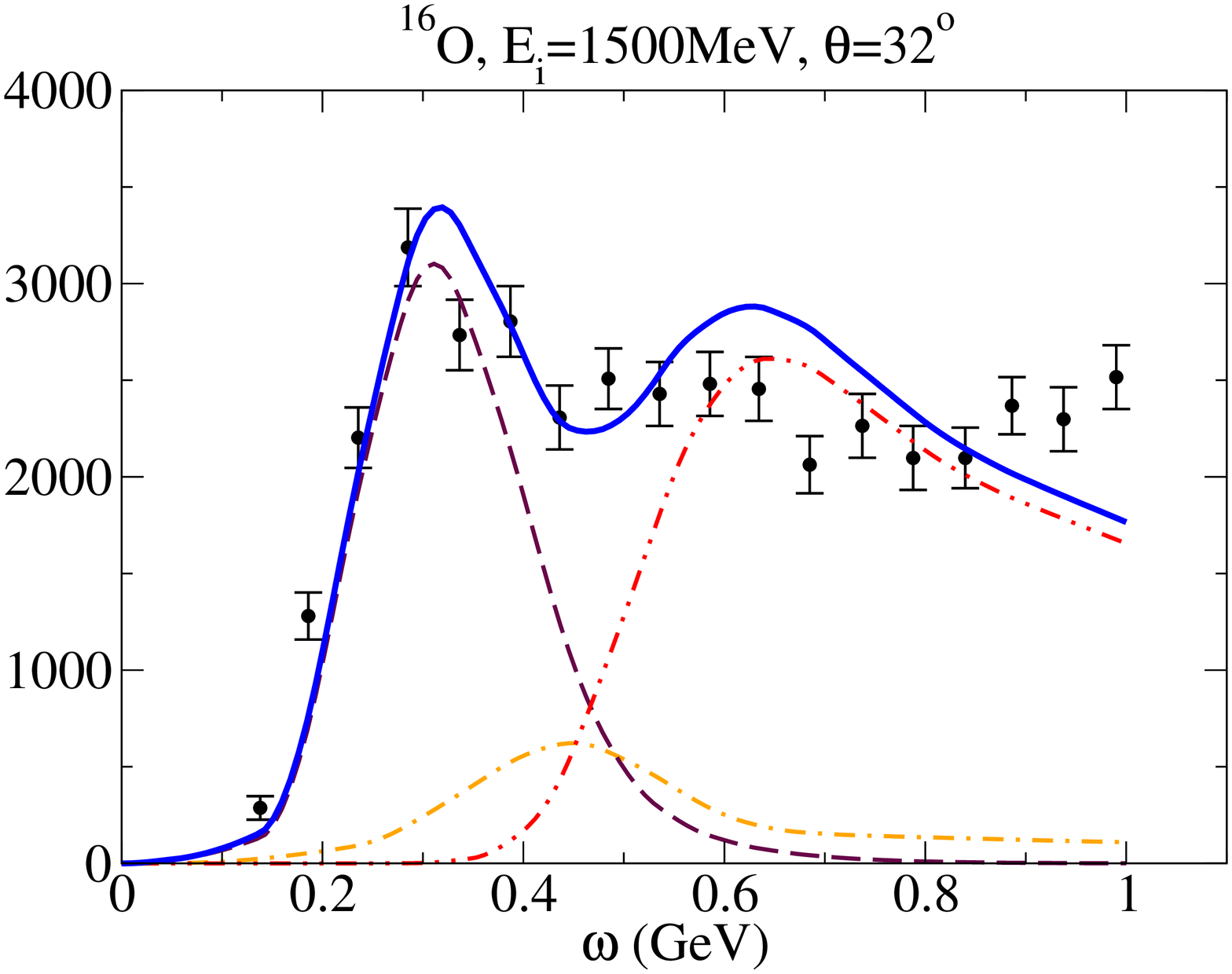}
\end{center}
\caption{\label{fig:fig1} (Color online) 
  Comparison of inclusive $^{16}$O$(e,e')$ cross sections and predictions of the SuSAv2-MEC model. The separate contributions of the pure QE response (dashed violet line), the 2p2h MEC (dot-dashed), inelastic (double-dot dashed) are displayed. The sum of the three contributions is represented with a solid blue line. 
  The data are from \cite{Anghinolfi:1996vm} and \cite{OConnell:1987kww}.
Figure from Ref.~\cite{Megias:2017cuh}.}
  \end{minipage}
\end{figure}

Before showing predictions for neutrino scattering on an oxygen target, we validate the model by comparing with electron scattering data, as shown in
Fig.~\ref{fig:fig1}.
In the case of $^{16}$O the available $(e,e')$ data cover only
a limited kinematic region corresponding to  six
different sets of the electron incident energy $E_i$ and scattering angle $\theta$~\cite{Anghinolfi:1996vm,OConnell:1987kww}.
In the calculation we have employed the Gari-Krumpelmann (GKex) model for the
elastic electromagnetic form factors~\cite{Gari:1984ia}, whereas the inelastic
structure functions are described making use of the Bosted and Christy
parametrization~\cite{Bosted:2007xd,Christy:2007ve}.~ The contribution of the 2p2h MEC is also
included in both the longitudinal and transverse channels, although the latter are largely dominant in the electromagnetic case.
The value of the Fermi momentum is fixed to $k_F=230$ MeV/c.
In all the cases we present the separate contributions for the QE, 2p2h MEC and inelastic regimes.
As observed, the SuSAv2-MEC predictions are in very good accordance
with data for all kinematical situations. Although the relative role
of the 2p2h-MEC effects is rather modest compared with the QE and
inelastic contributions, its maximum is located in the dip region between the QE
and inelastic peaks. This makes 2p2h-MEC essential in order to
describe successfully the behavior of $(e,e')$ data against the
transferred energy $\omega$: data in the dip region can only be
reproduced by adding MEC effects to the tails of the QE and inelastic
curves. Indeed, at the peak of the 2p2h response the three contributions are comparable in size.

\begin{figure}
  \begin{minipage}{\textwidth}
  \begin{center}
\includegraphics[width=4.9cm,angle=270]{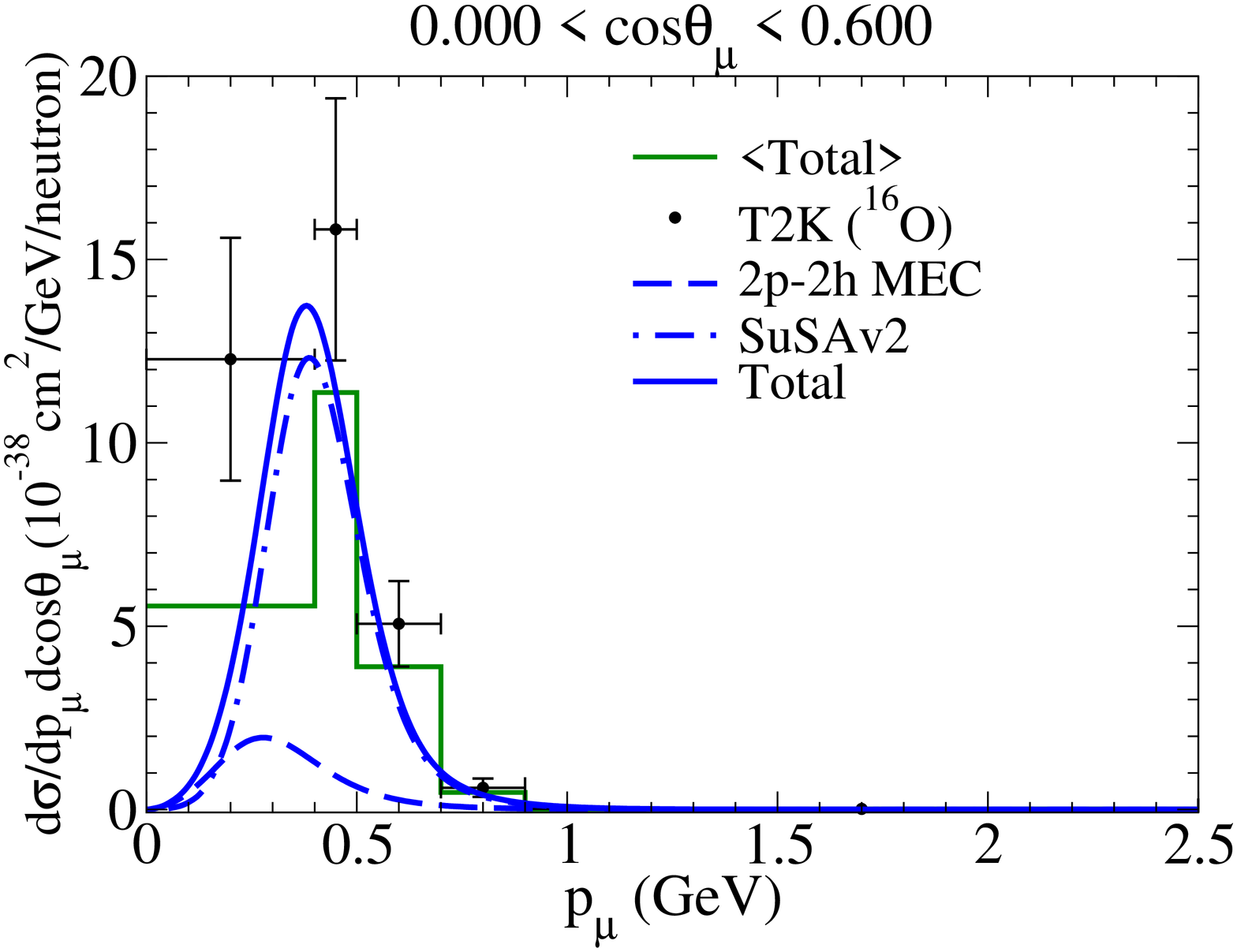}
\includegraphics[width=4.9cm,angle=270]{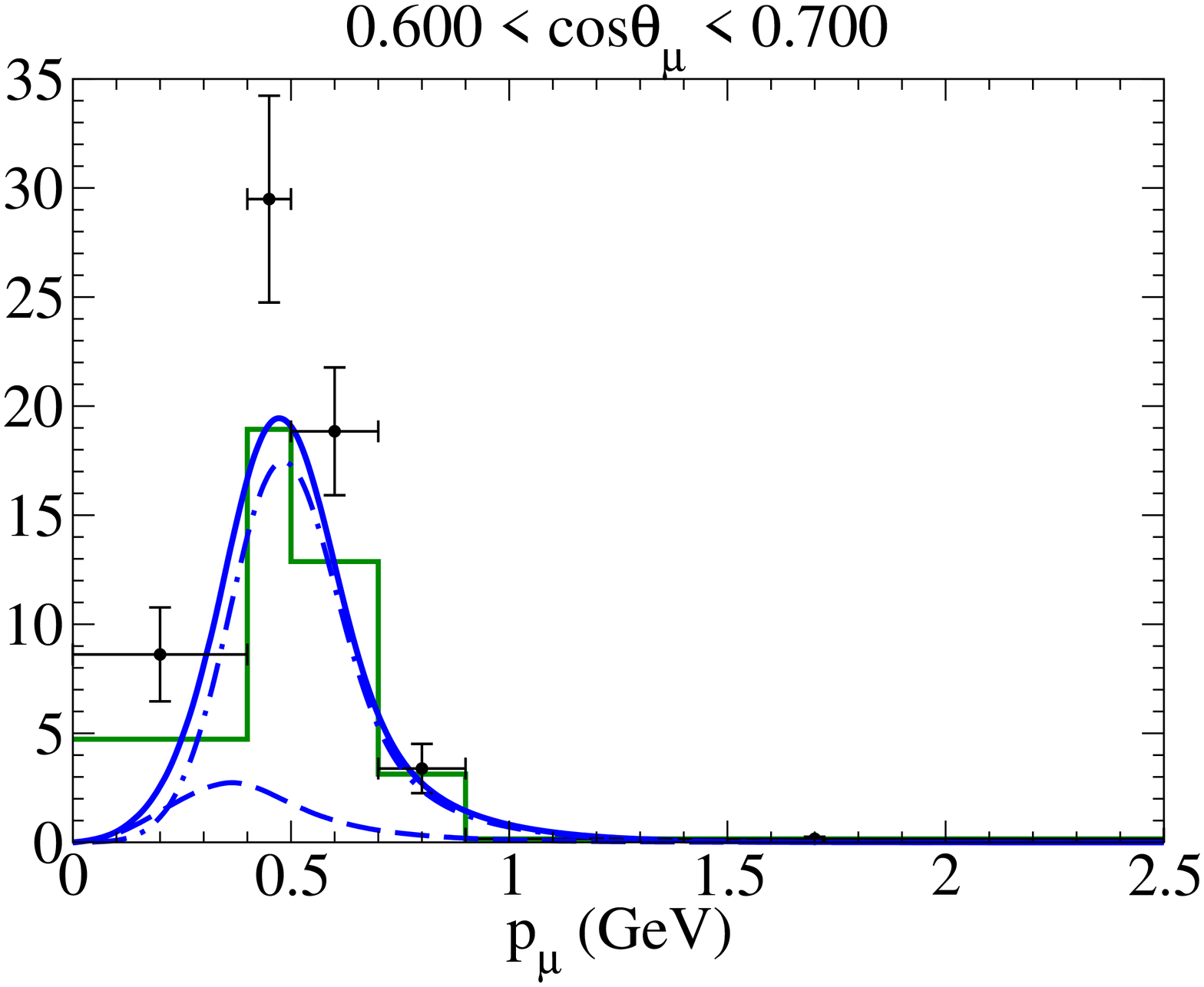}
\\
\includegraphics[width=4.9cm,angle=270]{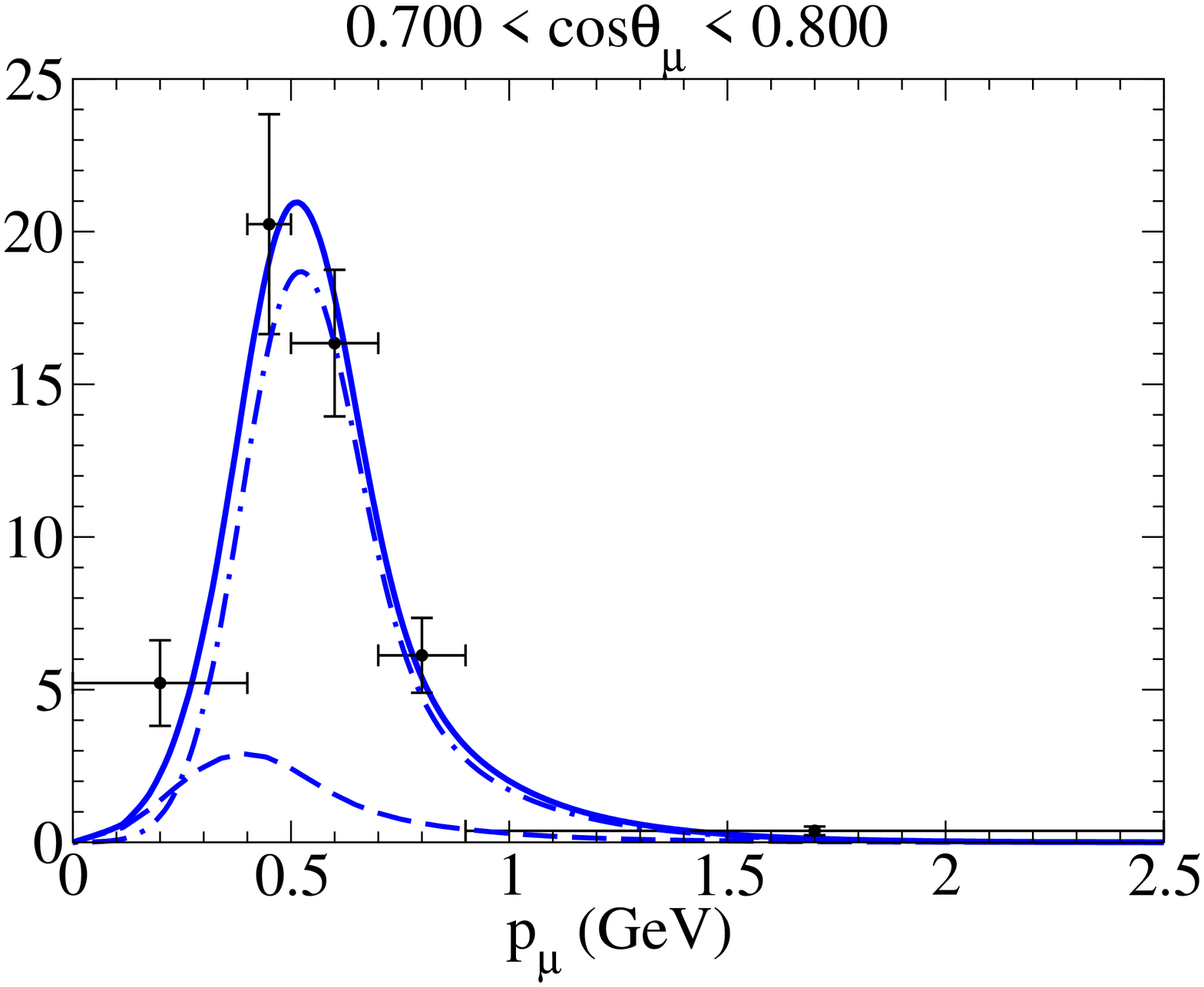}
\includegraphics[width=4.9cm,angle=270]{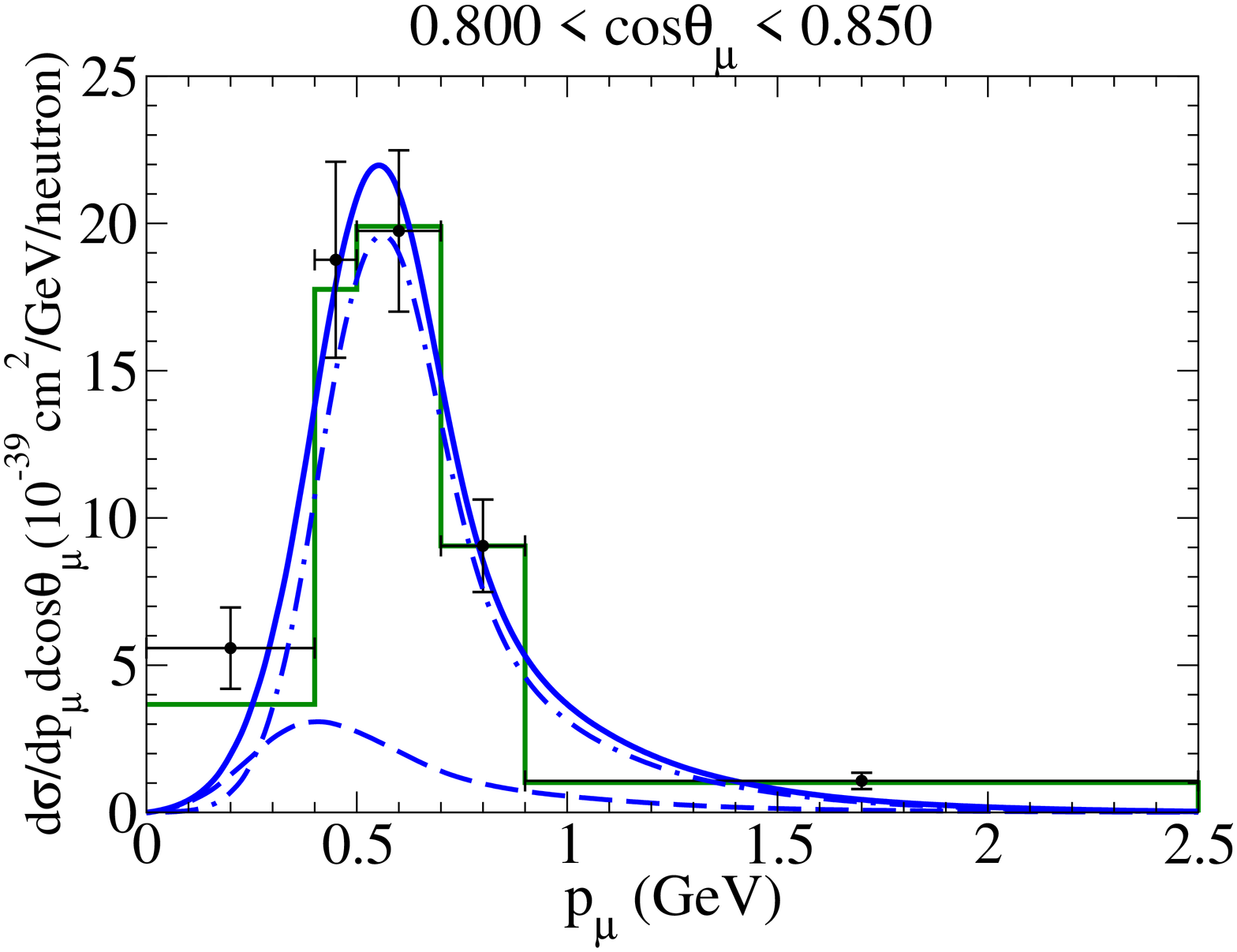}
\\
\includegraphics[width=4.9cm,angle=270]{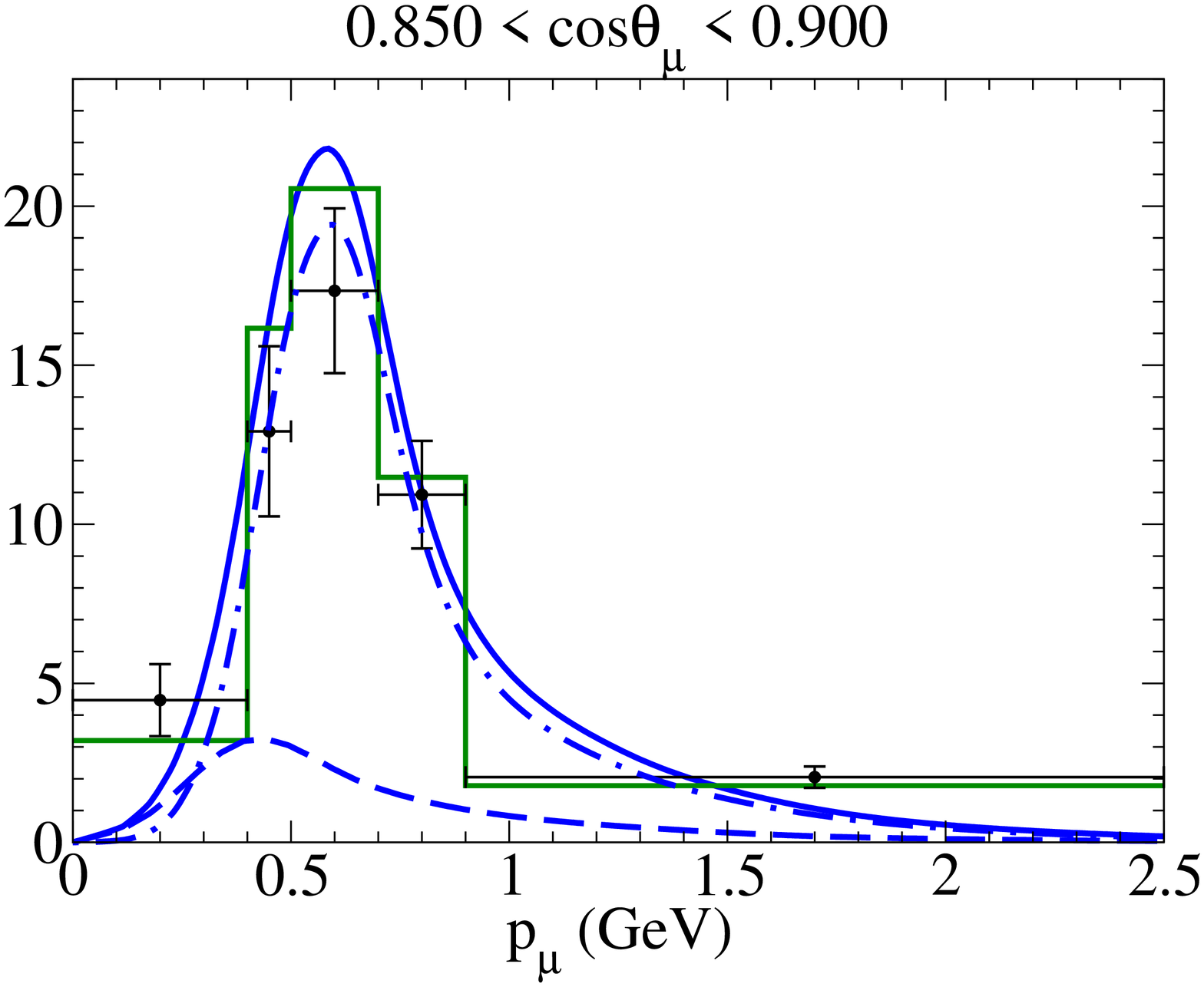}
\includegraphics[width=4.9cm,angle=270]{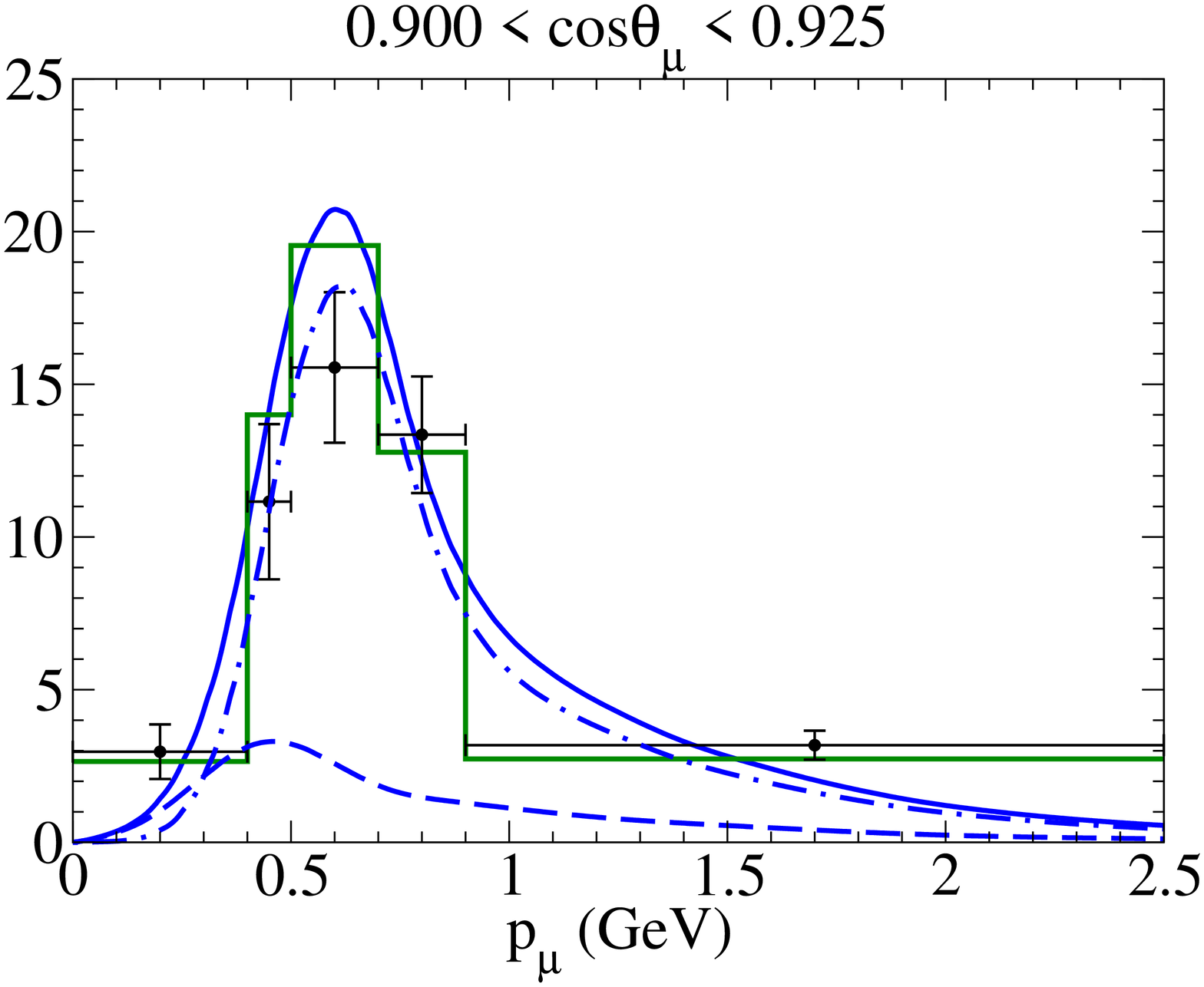}
\\
\includegraphics[width=4.9cm,angle=270]{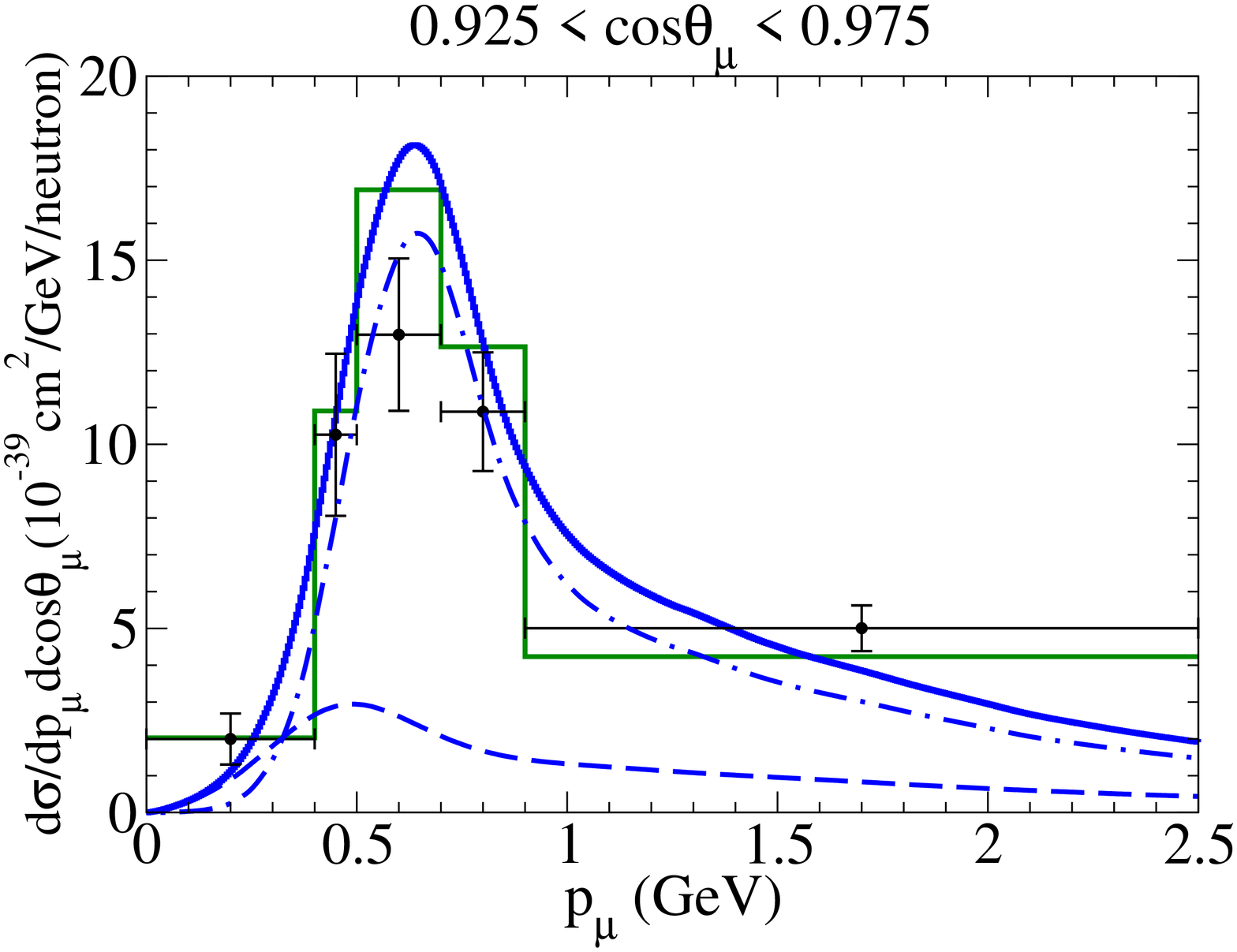}
\includegraphics[width=4.9cm,angle=270]{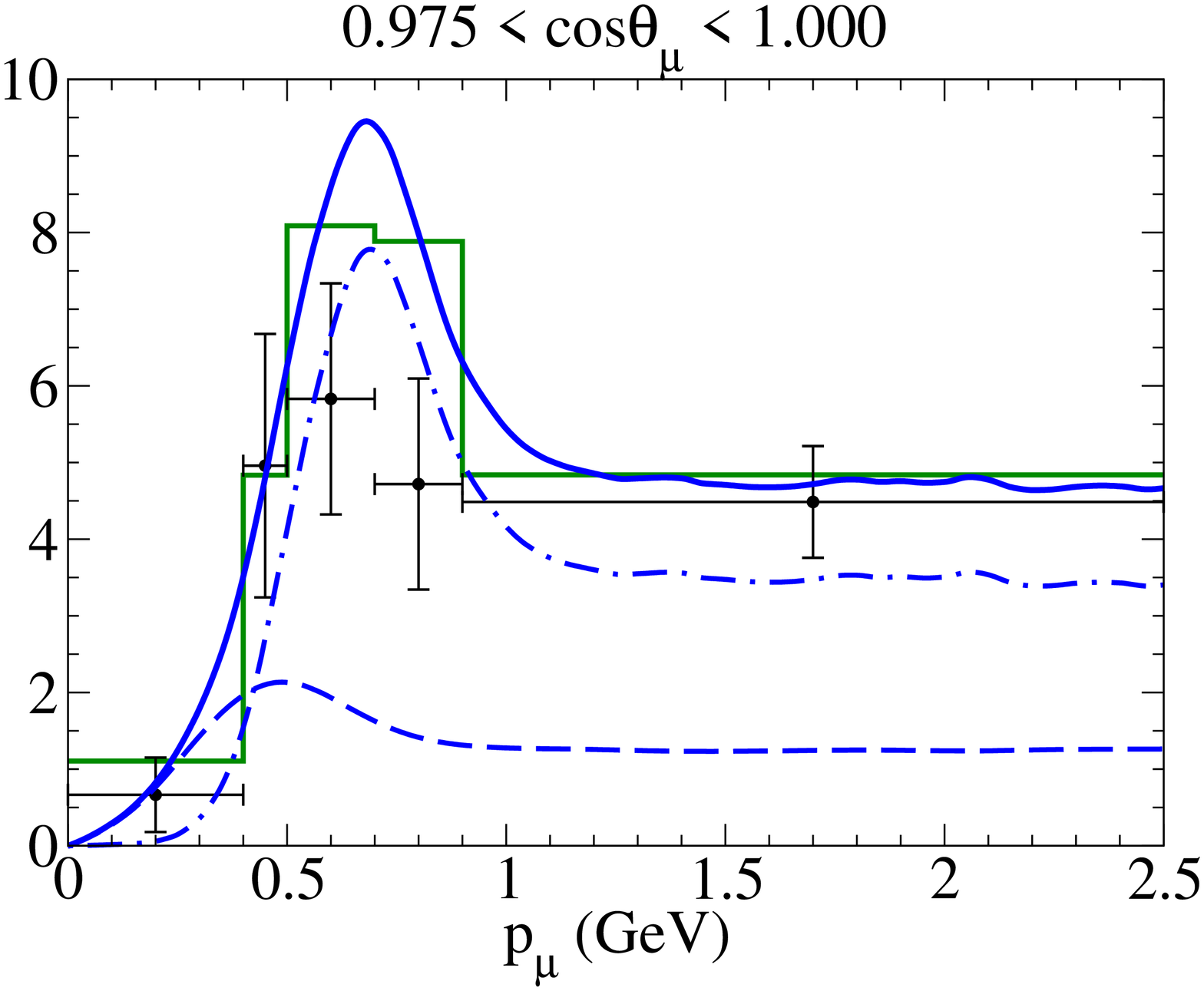}
\end{center}
\caption{\label{fig:fig2} (Color online)
T2K flux-folded double differential cross section per target neutron for the $\nu_\mu$ CCQE process on $^{16}$O displayed versus the
muon momentum $p_\mu$ for various bins of $\cos\theta_\mu$ obtained within the SuSAv2-MEC approach. QE and 2p2h MEC results
are shown separately. The histogram represents the theoretical average of the total result over each bin of $p_\mu$. The data are from~\cite{Abe:2017rfw}.
Figure from Ref.~\cite{Megias:2017cuh}.}
  \end{minipage}
\end{figure}

Having successfully tested the model, we now show in Fig.~\ref{fig:fig2} the results for CC neutrino reactions on $^{16}$O.
Each panel presents the double differential cross section averaged over the T2K muonic neutrino flux versus the muon momentum for fixed bins of the muon scattering angle. These kinematics
correspond to the recent $(\nu_\mu,\mu)$ CC0$\pi$ data collected by the T2K
experiment~\cite{Abe:2017rfw}.
Contrary to the $(e,e')$ cross sections previously shown, here only the QE and 2p2h MEC contributions are taken into account, as this is consistent with the analysis of the data, that is restricted to charged-current
processes with no pions in the final state .
We show the separate contributions of the pure QE, the 2p2h MEC and the sum of
both. Notice the role of the MEC effects compared with the pure
QE ones --- of the order of $\sim$15$\%$ at the maximum
of the peak, except for forward angles, where they represent about 20\% of the total cross section. Furthermore, the MEC peak compared with the QE one is
shifted to smaller $p_\mu$-values. These results, which are also
observed in the case of T2K-$^{12}$C (see \cite{Megias:2016fjk}), are
somehow different from the ones found in the analysis of other experiments, namely, MiniBooNE
and MINERvA, that show 2p2h MEC relative effects to be larger and the
peak location more in accordance with the QE maximum. This is
connected with the much narrower distribution presented by the T2K
neutrino flux, that explains the smaller 2p2h MEC contribution and the
location of its peak.  

The SuSAv2-MEC approach provides predictions in good agreement with T2K data in
most of the situations, although here 2p2h MEC effects do not seem to
improve in a significant way the comparison with data. This is at
variance with other experiments, MiniBooNE and MINERvA, and it is
connected with the minor role played by MEC. Notice that in most of
the situations, both the pure QE and the total QE+MEC predictions
describe data with equal success. A similar discussion was already
presented in \cite{Megias:2016fjk} for $^{12}$C.

The model predictions for antineutrino $\bar\nu_\mu$ scattering on water, for which data are not yet available, can be found in Ref.~\cite{Megias:2017cuh}.

\begin{figure}
  \begin{minipage}{\textwidth}
      \begin{center}
\includegraphics[width=8.4cm,angle=270]{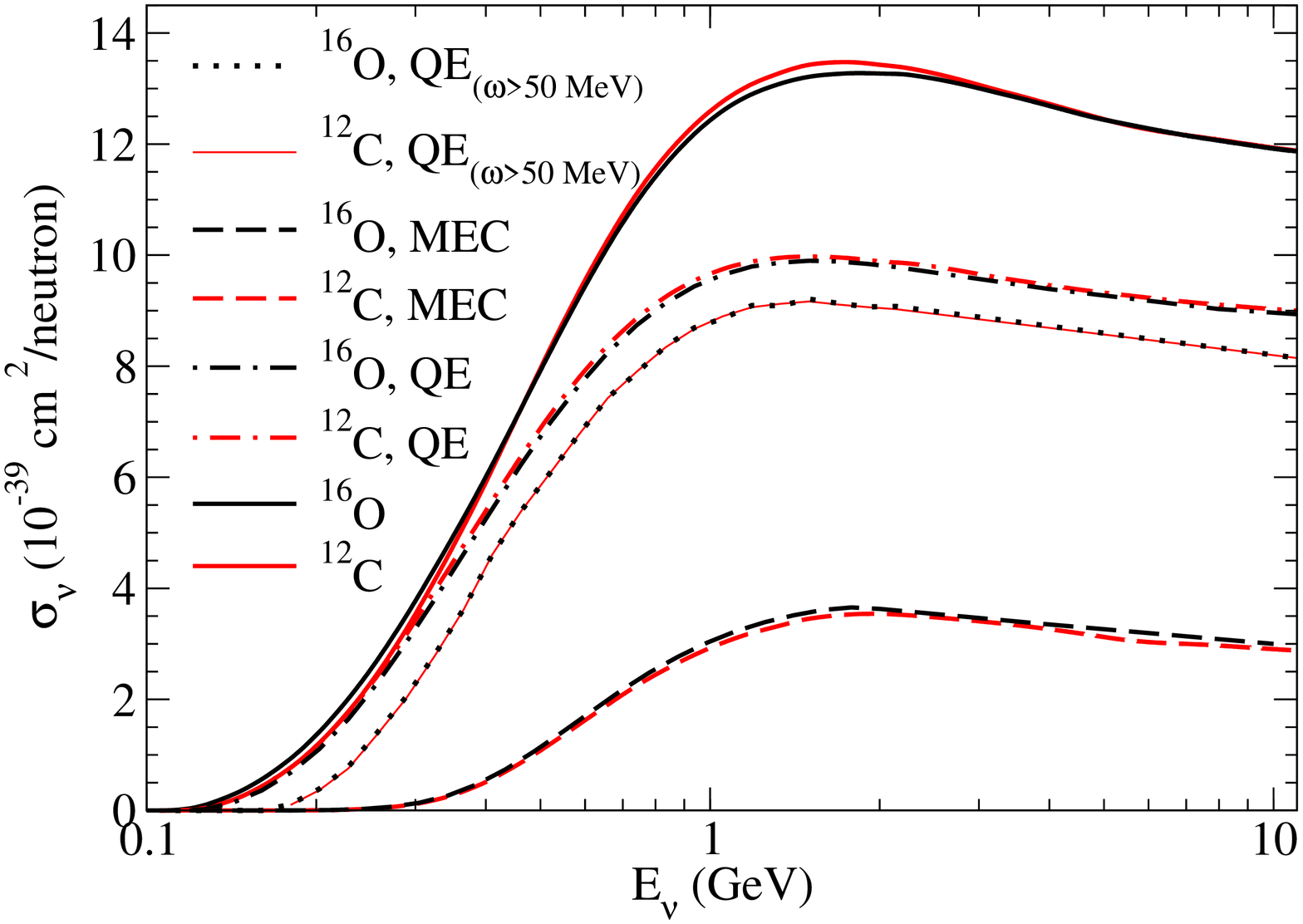}
\end{center}
\caption{\label{fig:fig4} (Color online) Total $\nu_\mu$
  cross section per nucleon as a function of the neutrino energy
  evaluated for $^{12}C$ and $^{16}O$ nuclei. Separate contributions of
  the pure QE (dot-dashed) and 2p2h MEC (dashed). The effect of making a cut in $\omega$ below 50 MeV for the QE contribution is also shown.
  Figure from Ref.~\cite{Megias:2017cuh}.}

  \end{minipage}
\end{figure}

Finally, we illustrate the dependence of the C/O differences upon the neutrino energy by displaying in Fig.~\ref{fig:fig4} the total integrated cross section per neutron with no neutrino flux included versus the neutrino energy. More detailed comparisons can be found in  Ref.~\cite{Megias:2017cuh}. The results shown here indicate that nuclear effects between these nuclei in the total cross section, that is, including both  the QE and 2p2h MEC contributions, are very tiny, at most of the  order of $\sim$2-3$\%$.  This minor  difference is also observed for the pure QE response (slightly higher for carbon) and the 2p2h MEC (larger for oxygen). This is connected with the  differing scaling behavior shown by the QE and 2p2h MEC responses  with the Fermi momentum, and the very close values of $k_F$ selected  for the two nuclei. Upon including both  the QE and 2p2h MEC contributions, one observes that nuclear effects in the total cross section are very tiny. We also show the effect of making a ``cut'' at $\omega = 50$ MeV, namely, setting any contribution from below this point to zero. This has been used in past work as a crude sensitivity test to ascertain the relative importance of the near-threshold region. If significant differences are observed when making the cut, then one should have some doubts about the ability of the present modeling (indeed, likely of all existing modeling) to successfully represent the cross section in this region. What we observe for the total cross section shown in Fig.~\ref{fig:fig4} are relatively modest effects from near-threshold contributions, although one should be aware that this is not so for differential cross sections at very forward angles where small-$\omega$ contributions can be relatively important (see \cite{Megias:2017cuh}).

Summarizing, the SuSAv2 model, based on superscaling and complemented with the addition of 2p2h contributions induced by meson exchange currents, has been applied to the simultaneous study of electron and CC neutrino scattering for two different nuclei, carbon and oxygen. Good agreement is found with all existing data. The scaling properties of the 2p2h response versus the nuclear density have also been analyzed and a scaling law, different from the one obeyed by the QE response, has been defined. This can be useful to estimate the importance of these contributions in different nuclei and extrapolate results from one nucleus to another without performing the explicit calculation. 
Given the success of the comparison of the model predictions with both inclusive $(e,e')$ and CC$\nu$ data, we have increased confidence in employing the approach for heavier nuclei. New features are likely to emerge in these cases and we shall explore their consequences in forthcoming work.

\end{document}